\documentclass[acmsmall,nonacm]{acmart}

\usepackage{amsmath} 

\AtBeginDocument{%
  }

\begin{document}

\title{Ultra Ethernet's Design Principles and Architectural Innovations}

%

\author{Torsten Hoefler}
\affiliation{%
  \institution{ETH Zurich, Switzerland \& Microsoft}
  \city{Zurich}
  \country{USA}}
\email{htor@ethz.ch}

\author{Karen Schramm}
\affiliation{%
  \institution{Broadcom}
  \city{San Jose}
  \country{USA}
}
\email{karen.schramm@broadcom.com}

\author{Eric Spada}
\affiliation{%
	\institution{Broadcom}
	\city{San Jose}
	\country{USA}
}
\email{eric.spada@broadcom.com}

\author{Keith Underwood}
\affiliation{%
	\institution{Hewlett Packard Enterprise}
	\city{San Jose}
	\country{USA}
}
\email{keith.underwood@hpe.com}

\author{Cedell Alexander}
\affiliation{%
	\institution{Broadcom}
	\city{San Jose}
	\country{USA}
}
\email{cedell.alexander@broadcom.com}

\author{Bob Alverson}
\affiliation{%
	\institution{Hewlett Packard Enterprise}
	\city{San Jose}
	\country{USA}
}
\email{bob.alverson@hpe.com}

\author{Paul Bottorff}
\affiliation{%
	\institution{Hewlett Packard Enterprise}
	\city{San Jose}
	\country{USA}
}
\email{paul.bottorff@hpe.com}

\author{Adrian Caulfield}
\affiliation{%
	\institution{OpenAI}
	\city{San Jose}
	\country{USA}
}
\email{adrianc@openai.com}

\author{Mark Handley}
\affiliation{%
	\institution{OpenAI}
	\city{San Jose}
	\country{USA}
}
\email{markh@openai.com}

\author{Cathy Huang}
\affiliation{%
	\institution{Intel}
	\city{San Jose}
	\country{USA}
}
\email{cathy.huang@intel.com}

\author{Costin Raiciu}
\affiliation{%
	\institution{Broadcom}
	\city{San Jose}
	\country{USA}
}
\email{costin.raiciu@broadcom.com}

\author{Abdul Kabbani}
\affiliation{%
	\institution{Microsoft}
	\city{Redmond}
	\country{USA}
}
\email{abdulkabbani@microsoft.com}

\author{Eugene Opsasnick}
\affiliation{%
	\institution{Broadcom}
	\city{San Jose}
	\country{USA}
}
\email{eugene.opsasnick@broadcom.com}

\author{Rong Pan}
\affiliation{%
	\institution{AMD}
	\city{San Jose}
	\country{USA}
}
\email{rong.pan@amd.com}

\author{Adee Ran}
\affiliation{%
	\institution{Cisco}
	\city{San Jose}
	\country{USA}
}
\email{aran@cisco.com}

\author{Rip Sohan}
\affiliation{%
	\institution{AMD}
	\city{San Jose}
	\country{USA}
}
\email{rip.sohan@amd.com}

\renewcommand{\shortauthors}{Hoefler et al.}

\begin{abstract}
The recently released Ultra Ethernet (UE) 1.0 specification defines a transformative High-Performance Ethernet standard for future Artificial Intelligence (AI) and High-Performance Computing (HPC) systems. This paper, written by the specification's authors, provides a high-level overview of UE's design, offering crucial motivations and scientific context to understand its innovations. While UE introduces advancements across the entire Ethernet stack, its standout contribution is the novel Ultra Ethernet Transport (UET), a potentially fully hardware-accelerated protocol engineered for reliable, fast, and efficient communication in extreme-scale systems. Unlike InfiniBand, the last major standardization effort in high-performance networking over two decades ago, UE leverages the expansive Ethernet ecosystem and the 1,000x gains in computational efficiency per moved bit to deliver a new era of high-performance networking.
\end{abstract}

\maketitle

\section{Introduction}
Ultra Ethernet (UE) standardizes a new protocol to support high-performance Artificial Intelligence (AI) and High-Performance Computing (HPC) networking over Ethernet. This paper, written by UE's authors, supplements the full specification by highlighting historical and innovative technical aspects of our nearly 2.5-year journey. It is designed to be approachable to a general audience and, thus, abstracts many details while using intuitive wording and explanations. The final authority for questions regarding UE is the full 562-page specification \cite{UltraEthernet1.0}.

In 2022, as the world was rapidly moving into the new age of massive computation required to satisfy the demands of artificial intelligence systems, various datacenter providers recognized the limitations posed by InfiniBand and its sister protocol Remote Direct Memory Access (RDMA) over Converged Ethernet (RoCE). At the same time, Ethernet's success as generic interconnect was hard to overstate with hundreds of millions of ports deployed per year and earning it's inventors the Turing award in that year. RoCE version 2 embedded InfiniBand's transport in a routable Ethernet (OSI Layer 3) to enable datacenter deployment nearly a decade earlier. RoCE adopted InfiniBand’s transport protocol largely unchanged, requiring lossless transport with strict in-order packet delivery. This lossless in-order packet delivery is guaranteed by converged Ethernet using priority flow control (PFC) as a main mechanism. Yet, PFC requires a separate traffic class with substantial headroom buffer and suffers from congestion spreading and head-of-line blocking. Furthermore, in-order delivery limits the choice of paths leading to potentially suboptimal performance. Those and further limitations of RoCE are summarized in \cite{roce-issues}. 

The original InfiniBand transport was designed more than a quarter century ago in an environment where architects had to exploit high bandwidth with limited computation~\cite{InfiniBand1.0}. After 25 years of exponential Moore’s Law scaling, the cost per transistor and, therefore, computation was reduced by more than 100,000 times while the usable bandwidth grew only by a factor of 100 from Single Data Rate (SDR) to eXtended Data Rate (XDR). Thus, not only did the accelerators at the endpoints get a significant performance boost, but also network architects have more than 1,000x more computation per moved bit. This prompted many to rethink the network stack in internal AI product lines~\cite{aws-efa,maia-patents,google-tpu} and HPC deployments \cite{slingshot}. Some early on recognized that datacenter and HPC networks are bound to converge into a single technology \cite{dc-convergence} and discussions among a handful of companies began to make this happen. 

The initial small group formed in the first quarter of 2022 including AMD, Broadcom, HPE, Intel, and Microsoft, who agreed to build an open standard for next generation Ethernet inspired by the various parallel internal developments and the opportunity to create a bigger market. Initially, the effort was called HiPER and later renamed to Ultra Ethernet (UE). By July 2022, a new consortium was formed, and we met face to face for the first time in September 2022. That first meeting would bring out a wide range of opinions. From “let’s standardize one of those RoCE variants” to “let’s build a new standard from the many ingredients we have”, the energy was high and the group was engaged. By January 2023, we had our path: we would combine capabilities designed for HPC and security and congestion management techniques designed for the datacenter to build something new: a highly scalable transport to run over plain old Ethernet. Shortly after, in July 2023, the Ultra Ethernet Consortium (UEC) was officially announced by AMD, Arista, Broadcom, Cisco, Eviden (Atos), HPE, Intel, Meta, and Microsoft. As an open Linux Foundation Joint Development Foundation (JDF) project, it quickly grew to more than 100 member companies than 1,500 participants at the end of 2024. 

UEC set out to define an open next-generation HPC and AI network compatible with existing Ethernet deployments and interoperable between vendors. The discussions focused on several foundational tenets:

\textbf{Massive scalability} enables large deployments that are needed for future AI systems. UE is designed to support millions of network endpoints in flexible arrangements with a connectionless API. Initially, it focuses on supporting traditional fat tree deployments, while not preventing other optimized topologies such as HammingMesh~\cite{hxmesh}, Dragonfly~\cite{df}, or Slim Fly~\cite{sf}, which the consortium did not test. 

\textbf{High-Performance} is achieved by efficient protocols designed for large-scale deployments. For example, UE’s connection-less API is supported by a mechanism to establish a peer-to-peer reliability context without additional latencies, that is, the initial packet arriving establishes a context, potentially in nanoseconds, even for massive out-of-order delivery. Furthermore, UE supports optional extensions such as packet trimming for fast packet loss detection and reaction.

\textbf{Compatibility} with existing Ethernet datacenter deployments is achieved by imposing minimal requirements on the switch infrastructure to allow easy deployment and rolling extension of existing infrastructures. UE switches are only required to support Equal-Cost Multi-Pathing (ECMP) and basic Explicit Congestion Notification (ECN) marked at egress, but they may (optionally) support packet trimming to improve network performance. UE does not require changes in either the Physical PHY Layer (OSI Layer 1) or the Link Layer (OSI Layer 2) but defines several optional extensions to improve the performance of the Link Layer in new deployments to foster vendor differentiation. UE is fully Ethernet-compatible, which enables users to use existing tools for operational management, debugging, and deployment. 

\textbf{Vendor differentiation} is maximized within the limits of interoperability of the specification. This enables the existing Ethernet vendor ecosystem to drive quick innovation cycles and development to its full extent in an active and sizeable market. In many cases, such as packet load balancing or fast loss detection, the specification proposes a set of options to implement compatible protocols but does not mandate any. Vendors could use one of the proposed approaches or invent their own to innovate. This enables architects to design their system to support important optimization objectives. For example, a vendor could choose specific load balancing and loss detection schemes to support the most reliable high-performance systems that are easy to operate. 

\subsection{Scope}

\begin{figure}[h!]
  \centering
  \includegraphics[width=\linewidth]{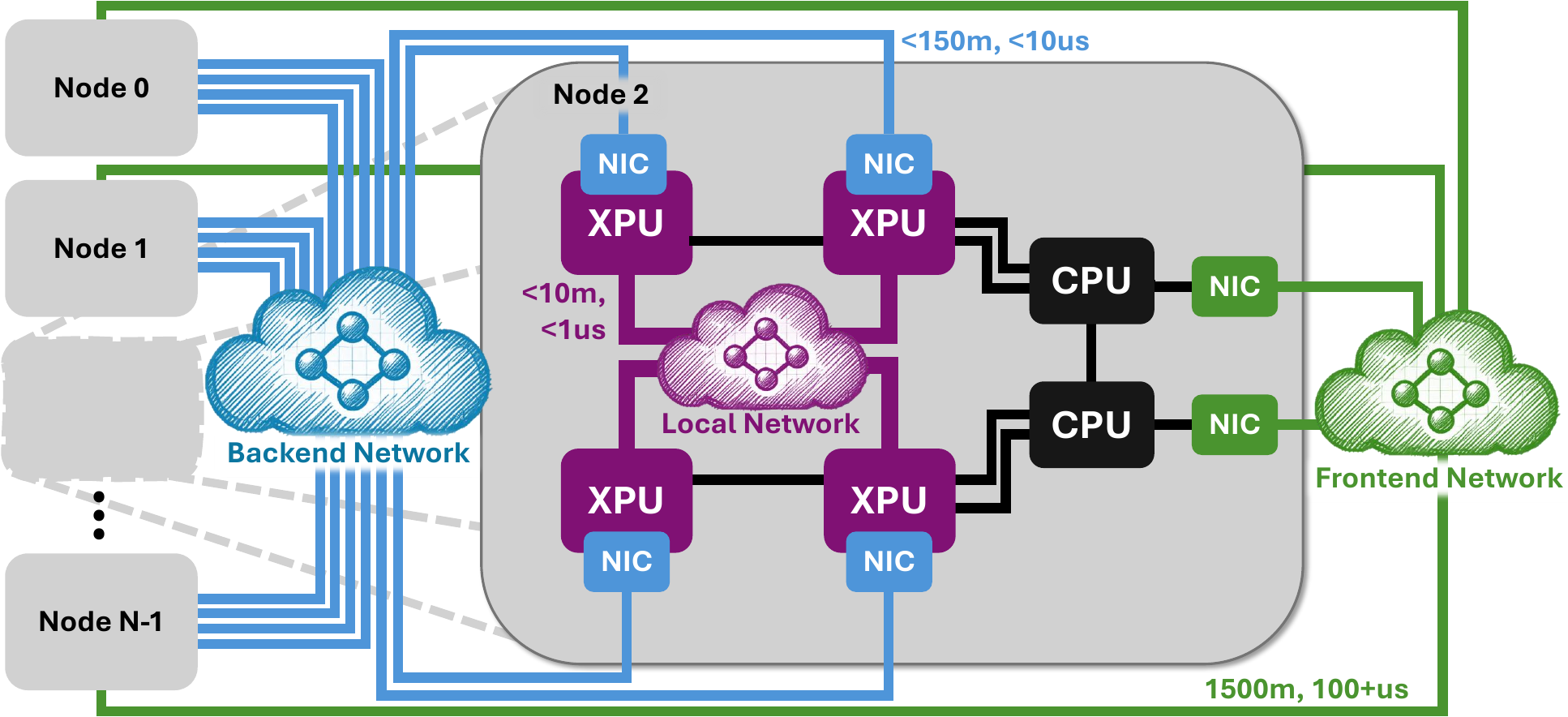}
  \caption{Overview of Network types in High-Performance Datacenters.}
  \label{fig:scope}
\end{figure}

Ultra Ethernet distinguished three fundamental network types: the local network (often called scale-up), the backend network (often called scale-out), and the frontend network. Figure~\ref{fig:scope} illustrates the three types of networks. The local network (purple) is used to connect CPUs and accelerators (XPUs) alike, today’s example deployments use CXL, NVLINK, or Ethernet for such node- or rack-scale networks with up to 10m range and sub-microsecond latency targets. The frontend network (green) is the traditional datacenter network carrying internal (“east-west”) as well as external (“north-south”) traffic. The backend network (blue) is the high-performance network that connects the compute devices (e.g., accelerators). Both backend and frontend networks are often called “scale-out” networks and they may be realized in a single physical instantiation. In fact, UE supports such a unification of those networks while also enabling deployments with physically separate instances. 

UE 1.0 aims mainly at the backend network. While it can be used for (shared) frontend or local networks, it is primarily designed for backend networks with the following assumptions: UE is designed to operate at high bandwidths (400+ Gbps) over medium-length (10-150m) links with relatively large messages and packets. Thus, header size and processing latency are important but not a primary concern. However, cheap high bandwidth and scalability to extreme systems are crucial. Designing for other network types, such as local or frontend would have emphasized other primary goals such as low-latency small-packet efficiency or simplest maintainability. Future versions of UE may target these network types more explicitly.

\section{Ultra Ethernet’s Key Features}

UE is designed to run seamlessly on existing Ethernet networks. The specification recommends running UE traffic in its own traffic class, but its congestion control algorithm is likely to work with other traffic, sharing the same switch buffers. It uses Layer 3 Internet Protocol (IPv4 or IPv6) compatible (routable) addressing and headers for seamless integration. UE defines Fabric Endpoints (FEPs) as logical entities that terminate the two ends of the transport layer for unicast operations. A FEP can be seen as roughly equivalent to a traditional Network Interface Controller (NIC).

The key features of Ultra Ethernet include:
\begin{enumerate}
\item		A highly scalable connectionless transport protocol using ephemeral Packet Delivery Contexts (PDCs).
\item	Removal of connection-oriented dependencies in the semantic layer, including buffer addressing, access authorization, and error models
\item	Native support for per-packet multipathing (“packet spraying”) with flexible and extensible load balancing schemes without reordering overhead at the receiver.  
\item	Both in-order and out-of-order reliable and unreliable packet delivery to optimally cover all use cases.
\item	Supports lossy (best effort) operation to avoid Head-of-Line blocking combined with optional packet trimming and other fast loss detection schemes for quick recovery.
\item	A novel congestion management scheme that adapts quickly to incast traffic and in-network congestion.
\item	A design that enables vendors to deliver hardware-only, software-only, or mixed hardware and software implementation products.
\item	Integrated scalable end-to-end encryption and authentication.
\item	Link-layer optimizations enabling accelerated implementations.  
\end{enumerate}
We will describe these and other features in detail in the following sections on the UE architecture. Before we continue, we describe ECMP-based packet spraying as a foundational concept for load balancing in UE.

\subsection{ECMP Packet Spraying}

Equal-Cost Multi-Pathing is a scheme for load balancing flows in the network. Switches that support ECMP will not resolve a target address directly to a port but to a set of ports that have similar-cost paths to the destination. Then, a deterministic hash function $p=H(x)$ is used to select the output port $p$ for each packet. The input to the hash function is often configurable and usually includes the full IP five-tuple (source and destination addresses and ports, as well as protocol type). Thus, if used without change, ECMP routes all packets of the same flow along the same deterministic path (in the absence of failures). UE redefines one of these fields to contain a so called Entropy Value (EV). For example, if standard UDP/IP is used, this field is the UDP source port, which is otherwise unused.  The UE Transport protocol (UET) has been assigned to UDP port 4793 by IANA. This is not only a beautiful large prime number, but also (++RoCEv2)++. UE also supports a native IP-only mode, where the EV is kept at the same position as the UDP source port. The source Fabric Endpoint (FEP) can now select a different EV for each packet that shall be sent on a different path or it can select in-order delivery when sending packets with the same EV. 

\begin{figure}[h!]
	\centering
    \includegraphics[width=\linewidth]{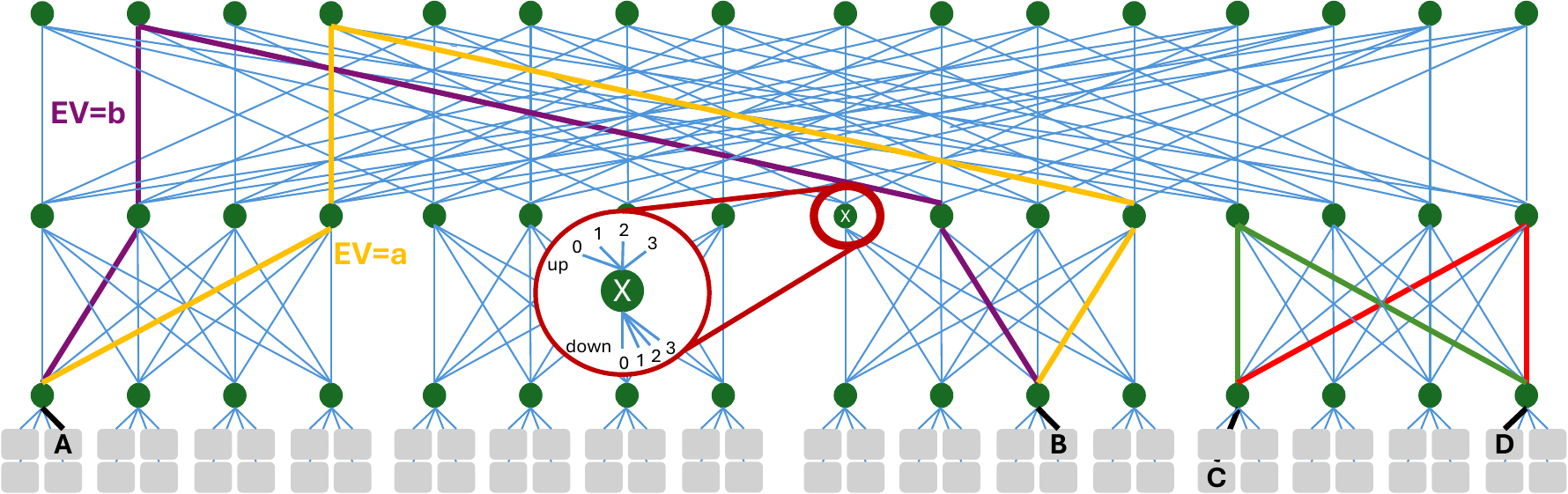}
	\caption{Entropy Values in Equal Cost Multi Pathing Illustrated.}
	\label{fig:ecmp}
\end{figure}

Figure~\ref{fig:ecmp} illustrates this scheme with a full Clos network constructed from 8 port switches (green circles) that support 64 endpoints (gray squares). We highlight Switch X in the second layer. It has four up and four down ports. Each packet that traverses a path first enters a switch through a down-facing port and leaves through the up-facing port unless the destination is an ancestor in the tree starting at the switch. Once the common ancestor switch between source and destination is reached in a Clos network, the packet turns towards a unique down path. If we consider the network shown with 64 endpoints in 4 groups of 16 endpoints each, then there are exactly four equal-cost (three switch hops) paths between any two nodes in the same group (cf. the green and red paths between nodes C and D). Furthermore, there are 16 paths between any two nodes in different groups (cf. the purple and yellow paths between nodes A and B). 

Due to its simplicity, ECMP is somewhat limiting because a node cannot directly select a path. Instead, it only knows that two packets with the same EV are guaranteed to take the same path (in a failure-free scenario). However, it does not know whether two different EV values may share the same path due to hash collisions. In fact, such hash collisions and related (sub)path sharing are expected! If we stay in the same group, there are only four different paths while there are $2^{16}$ EVs. Thus, for a perfectly distributing hash function, we expect a conflict probability of 25\% when selecting two random EVs. For nodes in different groups, this probability is still 6.25\%. Practical deployments use larger-radix switches where those probabilities may be different.

If two paths conflict, the bandwidth for each path is halved, and the performance loss can be detrimental. This is especially serious if the paths never change as in traditional Ethernet systems, which leads to a phenomenon called 'traffic polarization' \cite{b4}. 

UE's packet spraying can avoid such polarization by changing the EV for each packet, and thus, in expectation, distribute the packets evenly across all switches. Even if hash collisions occur, they will be short, and the resulting imbalance can be absorbed in the switch buffer. This leads to full network utilization and even traffic distribution on average over time. Packet spraying is simple if all endpoints spray evenly but becomes more challenging if some flows require in-order delivery and thus occupy some paths deterministically. UE proposes various optional load-balancing algorithms to determine how to set the EV for each packet. Finding the best such scheme remains open for vendor differentiation and research. 

\subsection{Ultra Ethernet Profiles}

The UE specification offers three profiles (HPC, AI Full, and AI Base) to support different feature sets allowing for implementations of varying complexity. The HPC profile offers the richest set of features, including wildcard tag matching, and is optimized for MPI and OpenSHMEM workloads. The AI Full profile is a superset of the AI Base profile. Both are aimed at Collective Communication Libraries (*CCLs) that do not require wildcard tag matching or other advanced communication operations. Both AI profiles offer deferrable sends, a feature specifically designed for *CCL communication offload. In addition, AI Full offers exact tag matching. AI Base is designed for the simplest implementation complexity, assuming that matching and other parts of the protocol are handled in the libfabric provider or the client library (e.g., a CCL). 

The HPC profile is a superset of the AI Base profile, and by implementing deferrable send, an implementation can provide both the HPC profile and AI Full profile.  The libfabric definition over UE enables an endpoint to advertise multiple profiles and then negotiate to the profile with the greatest common feature set. 

\section{Ultra Ethernet Architecture}

\begin{figure}[h!]
	\centering
    \includegraphics[width=\linewidth]{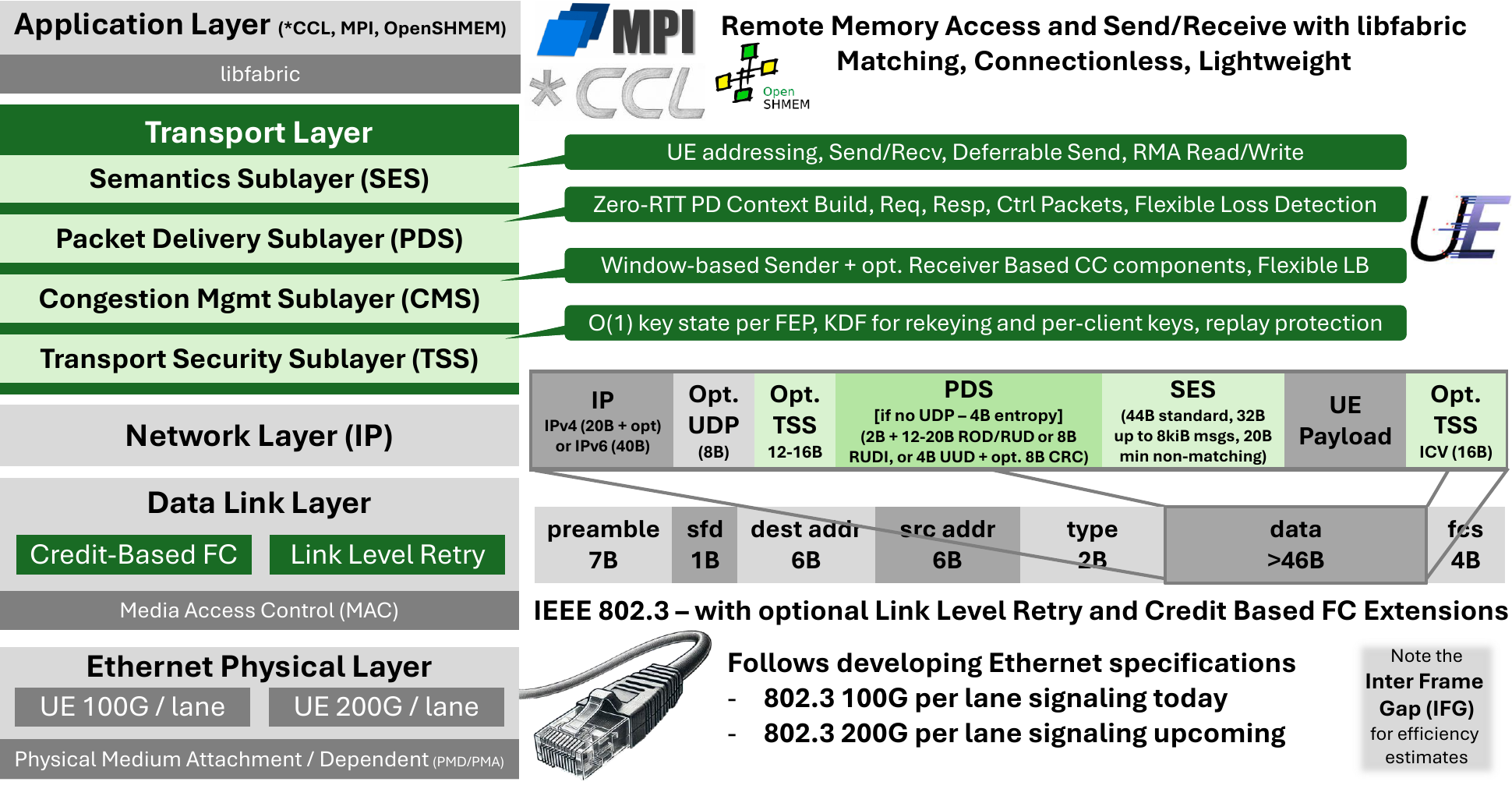}
	\caption{Ultra Ethernet's Overall Layered Architecture.}
	\label{fig:arch}
\end{figure}

We now dive into the components of the UE architecture. We structure this along the standard TCP/IP layering ranging from Layer 1 to Layer 4, as illustrated in Figure~\ref{fig:arch}. 

We start at the bottom left of the figure, the lowest layer in Ethernet, the \textbf{Physical (PHY) Layer}. This layer is essentially unchanged by UE to remain compatible with any Ethernet deployment. The first UE products will support 100G / lane signaling or 200G / lane signaling. The next layer is the \textbf{(Data) Link Layer}, which is also compatible with standard Ethernet to enable interoperability. It includes two optional and compatible UE extensions: Credit-Based Flow Control (CBFC) and Link Level Retry (LLR) that we will describe in Section~\ref{sec:ll}. The \textbf{Network Layer} utilizes standard IPv4 or IPv6 defined in the corresponding RFCs in order to allow legacy datacenter routing and operations. Within the network layer, switches may implement packet trimming, an optional feature that enables the destination to detect dropped packets in the network. The \textbf{Transport Layer} is by far the most significant change, as it is defined specifically for UE. It is designed to run over standard IP/UDP or natively over IP. It can be broken into four Sublayers: \textbf{Semantics (SES), Packet Delivery (PDS), Congestion Management (CMS), and Transport Security (TSS).}

The TSS encrypts the PDS and SES (including the payload) and authenticates the IP addresses to limit attacks that exploit the spoofing or 'guessing' of the transport header field. In conjunction with the PDS, the TSS aims to prevent replay attacks, and it is integrated with the semantics sublayer (SES) to reduce protocol layer attacks such as the misuse of credentials. The CMS works at the Byte level and controls the size of the outgoing window using congestion control contexts (CCCs). The PDS associates one or more Packet Delivery Contexts (PDCs) with each CCC and works at the packet level. PDS manages the reliable transmission of packets and cooperates with the SES with respect to message handling. The SES manages message transactions that are implemented using packets. It directly links to the libfabric interface in the application layer.  UE 1.0 uses libfabric as the interface to higher-level software and libraries such as CCLs and MPI. The SES also manages the execution of operations at the target, such as committing RMA operations into memory. 

\subsection{Transport Semantics Sublayer (SES)}

The libfabric interface utilizes the SES of the UE transport (UET) layer to provide the user-tagged send/receive and RMA operations. UE’s SES defines a wire protocol and semantics that are heavily inspired by the Portals 4 specification~\cite{portals-4.3} to enable an efficient and lightweight libfabric provider implementation. Unlike many network APIs, libfabric, like Portals 4, separates the traditional networking semantics – addressing, completion, authorization, and failure handling – from the notion of a connection. The way in which this difference drives the definition of the transport layer is explored in more detail below.

\subsubsection{Addressing}

Like most network APIs over Ethernet, UE uses IP addresses (the Fabric Address, or FA) to select endpoints (FEPs). To achieve scalable addressing, UE then uses JobIDs (24b), process identifiers at the destination FEP (12b PIDonFEP), and a Resource Index (12b RI) to address a logical context at each endpoint. UE defines two addressing modes that differ in the interpretation of PIDonFEP, which identifies a target address space (process) at each FEP. Finally, the resource index selects a receiver context in the target process, for example, a queue for send/receive or matching buffers and completion queues for RMA.

\begin{figure}[h!]
	\centering
    \includegraphics[width=\linewidth]{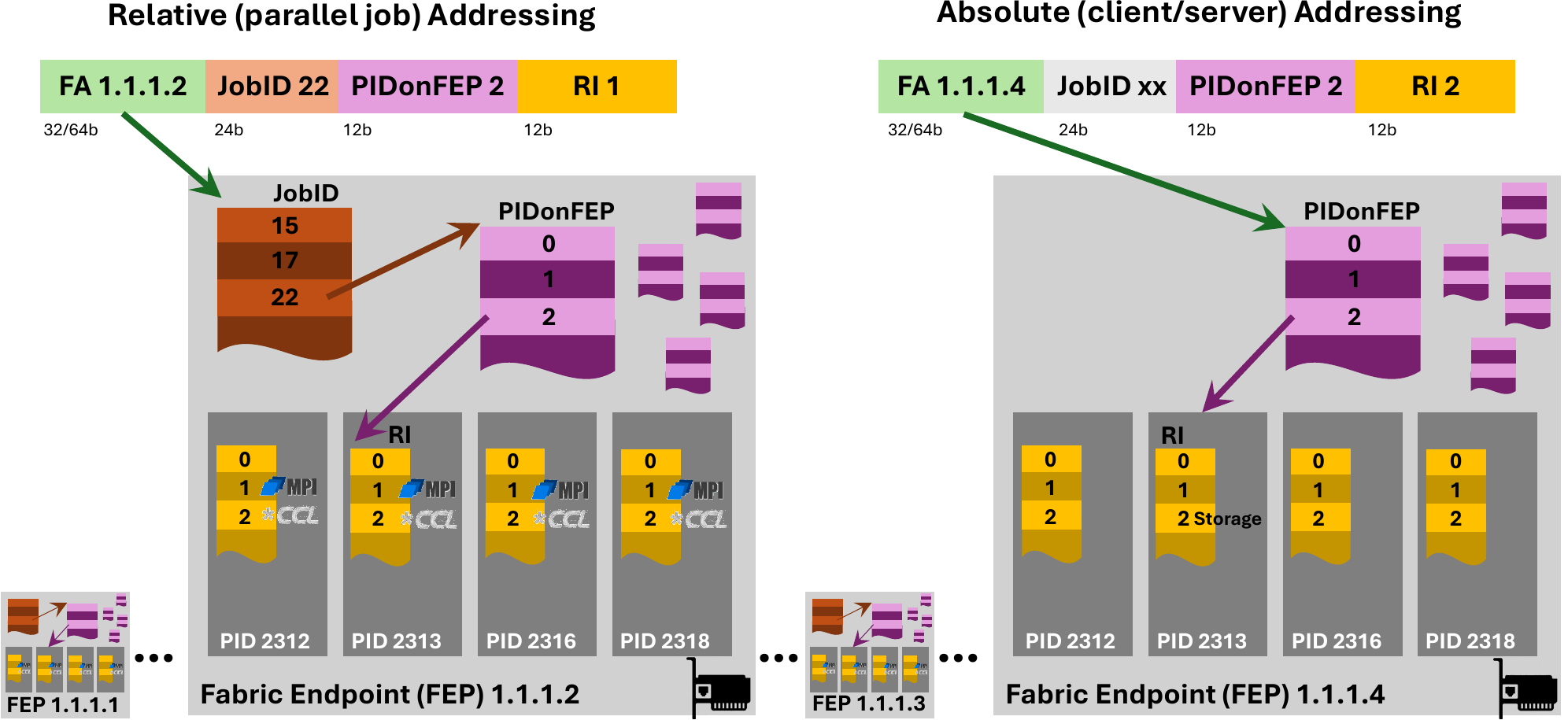}
	\caption{Ultra Ethernet Transport's Addressing Scheme.}
	\label{fig:addr}
\end{figure}

The two supported process addressing modes are \emph{relative addressing} for distributed parallel jobs in a cluster and \emph{absolute addressing} for client-server applications. They are distinguished by the \emph{rel} bit in the packet header. In relative addressing, each parallel job is identified by a unique JobID. Here, the FA identifies a FEP (gray node) in the system. Each FEP has a global JobID table which the incoming packet is matched to. Each JobID table entry identifies one job in the system that has processes running on this node, and it points to the local PIDonFEP table. The PIDonFEP table identifies all processes (address spaces) that are within the job identified by JobID. The table entry then resolves to a resource index table per process, which itself has services or other process-local resources attached. In the example shown in Figure~\ref{fig:addr}, the address identifies the MPI endpoint in Process 2313 on the node with FEP FA 1.1.1.2. The relative addressing scheme supports decoupling node-local and global addressing if all endpoints have the same number of processes. With direct addressing, each source process would need to store N*P many entries to address N nodes with P processes each. With relative UE addressing, each source node only stores N entries and can compute the target process as the PIDonFEP table offset relative to each node.

Absolute addressing, as shown in the right part of Figure~\ref{fig:addr}, is used for client/server applications to reach fixed network services such as storage or microservices. Such services are not part of jobs, and thus, the JobID is not used for addressing (it can be used as authentication token as it is still carried in the header). In absolute addressing mode, PIDonFEP can be used like a UDP port. UE also supports merging both PIDonFEP and Resource Index (RI) into a single table. 

In general, send/receive, tagged operations, and RMA operations use different RI tables. RMA operations carry an additional memory key (in the 64b match bits) that is used within an RI context to identify a target buffer. This memory key can be omitted in an optimized header format if buffers can be directly associated with RIs. Authorization is ensured through the JobID, which thus needs to be assigned by a trusted entity. In-network manipulation can be prevented using the TSS, described in Section~\ref{sec:tss}.

These addressing mechanisms enable one of the key scalability enhancements provided by UE. We illustrate it in contrast to existing mechanisms: InfiniBand’s Verbs~\cite{ibverbs}, for example, was originally designed to associate a receive queue with each queue pair (QP). However, users quickly learned that the memory required to associate a receive queue with every communication peer on a system with 10,000,000 cores was untenable; thus, shared receive queues were created. In contrast, the UE model allows users to address a queue without a context at the source. The JobID + PIDonFEP + RI uniquely addresses a single queue at a FEP. This makes the concept of shared received queues in traditional RDMA networks superfluous. Anyone in the application can send to this queue, and authorization to write to the queue is provided by the Job ID. 

A favorable side-effect of the design without queue-pairs acting as connections is that it enables simpler per transaction failures and error handling instead of the queue error states that add significant complexity in traditional RDMA systems.

\subsubsection{Messaging and Matching}

At the receiver, each RI has an associated receive queue and arriving messages match an entry in this queue. Untagged operations use entries in the order in which they were added at the receiver. Tagged messages select an entry using the tag. If the message is 'unexpected' (there is no entry), the receiver can discard the message and send a ‘buffer not ready’ response or buffer its headers. Optionally, the receiver may also buffer part of the payload until the receive buffer is posted, as described in Deferrable Send and Rendezvous below.
 
Hardware message matching is supported through a packet-carried initiator ID (32b) as well as a matching key (64b). An MPI or Collective Communication Library (CCL) program would encode the source rank in the initiator ID. The additional match key bits may be used to encode a communication context (e.g., an MPI communicator) and a message tag. The HPC profile supports in-order wildcard matching, which poses some implementation challenges in hardware~\cite{exampi-matching}. The AI Full profile uses exact matching without ordering to support simple hash-based or content-addressable memory (CAM) implementations. The AI Base profile does not support matching at the transport layer.

Why did we choose to incorporate matching? Simply put, matching is a fundamental semantic in HPC (e.g., MPI) that became an obviously useful tool for supporting the most common CCL messaging semantics over an unordered network. By constraining the AI Full profile to exact matching and simplifying unexpected message semantics (see below), the implementation can remain small.  In the process we gained a powerful tool: using tagged matching, messages (e.g., nccl\_send) can cross the network using an unordered protocol and, yet, arrive in the correct buffer.  This is accomplished by having the upper layer (i.e., the CCL) place a message sequence number into the match bits. 

\subsubsection{Handling Large Unexpected Messages: Deferrable Send and Rendezvous}

Historically, short, unexpected messages were simply buffered at the receiver; however, large, unexpected messages needed a different mechanism. The three profiles have different ways of handling large unexpected messages. A \textbf{large message} in UE is a message that exceeds the current send window size. In many existing communication libraries, it is fixed for simplicity, often defined by a configurable parameter called the 'eager limit' $s_e$. A message is called \textbf{expected} at the receiver if the receive address is known when the message arrives, that is, the receive has already been called by the application. It is called \textbf{unexpected} otherwise. Unexpected messages are often inevitable due to a variety of factors, such as load imbalance, and even are frighteningly common in some applications \cite{unexpected}.

\begin{figure}[h!]
	\centering
    \includegraphics[width=\linewidth]{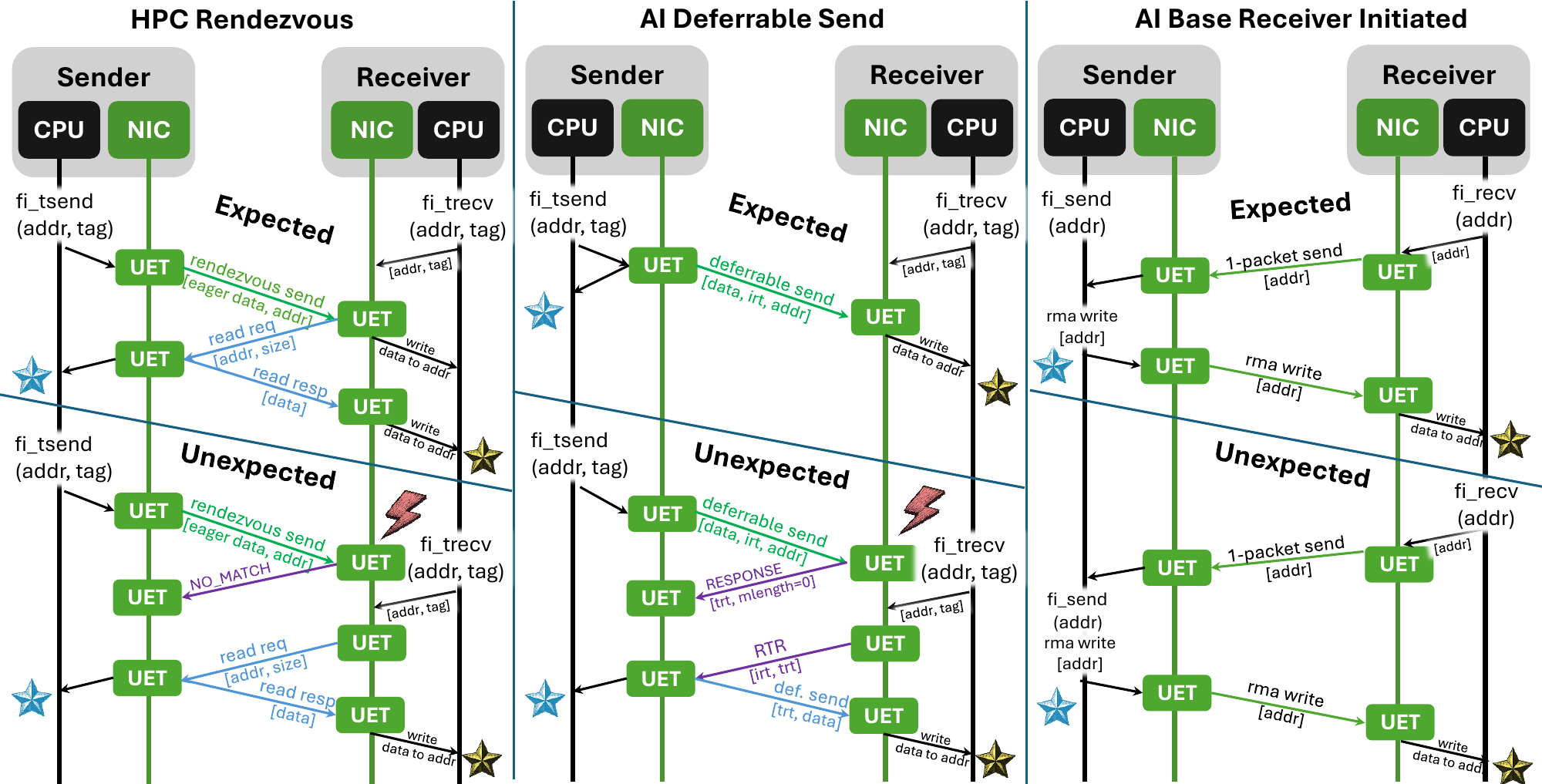}
	\caption{Ultra Ethernet Transport's Large-Message Send Protocols.}
	\label{fig:protos}
\end{figure}

\paragraph{Rendezvous}
In the traditional HPC rendezvous protocol, small messages are sent in a single step and large messages are sent in two steps – a first part of size $s_e$ together with a local addr of the remaining data followed by a second part retrieved via RMA read from the source addr. The left part of Figure~\ref{fig:protos} illustrates the protocol for both an expected and an unexpected large message. The implementation may query the current window size shortly before sending the rendezvous send to adjust the send size to the optimal value. The expected case shows the eager send step in green and the read step in blue. The unexpected case has an additional control message that informs the source that the message was not matched. 

\paragraph{Deferrable Send}
The deferrable send protocol is shown in the middle part of Figure~\ref{fig:protos}. An expected message is simply handled as a normal send, the message is sent in full and copied into the receive buffer, irrespective of the size. Small, unexpected messages may be buffered at the receiver, like in the rendezvous case. If the first packet of a large, unexpected message that cannot be buffered arrives, the receiver immediately replies with a response message that tells the source to defer (stop) sending. Large, unexpected messages allow for partial buffering of the packets (e.g., one window size) at the receiver. After the matching receive for a deferred message is posted at the receiver, a request to resume message is sent to the sender. To simplify the implementation, each deferrable send carries a token from the source that identifies the message (initiator restart token, irt), and each response carries the size of the locally stored data as well as a message identification token at the receiver (target restart token, trt). This use of tokens allows for simple table-based matching on both sides. Deferrable sends support an efficient low-resource hardware-offloaded implementation as they do not require RMA read support\footnote{Note that the UE consortium decided to make RMA read mandatory in AI Full to support storage use-cases.}. They can also be seen as an optimization of the traditional, sometimes painful, Receiver-Not-Ready (RNR) flow~\cite{InfiniBand1.0} in that the hard-to-configure timeout is replaced with an explicit signal. Furthermore, deferrable sends can react to changes in the size of the send window during sending and will therefore always send the optimal size to utilize the full bandwidth and avoid the well-known eager-to-rendezvous bandwidth drop~\cite{rendezvous}. 

\paragraph{Receiver Initiated} 
The AI Base profile does not require you to support any of these options. Instead, implementors can choose a receiver-initiated protocol driven by software, as shown in the right part of Figure~\ref{fig:protos}. Here, hardware support can be minimal and only support single packet send operations into specialized buffers together with RMA write only. The single packet send operation is used to send all information needed for the RMA write of all payload to the sender, which causes the sender to initiate the write from software (i.e., the provider implementation)~\cite{maia-patents}. The cost is that the sender needs some asynchronous activity (e.g., a thread) to initiate the write. Furthermore, the protocol adds one round-trip time (RTT) in the worst case when the send is posted exactly $\frac{1}{2}$ RTT before the receive is posted. 

In Figure~\ref{fig:protos}, UET boxes illustrate the participation of the UE transport layer in the flow. Red lightning indicates the arrival of an unexpected message. We assume that only headers are stored at the receiver (UE allows either to store nothing, which would require the sender to retransmit after a timeout, to store headers only, or to store headers and (partial) payload of the original transaction, which would lead to a local copy step upon posting the receive). Blue stars mark completion at the sender and yellow stars at the receiver. PDS acknowledgments are omitted for readability; if they were considered, the completion events at the source would be delayed by up to 1 RTT. 

We can now distinguish four different cases of all the combinations of expected, unexpected, small, and large for each of the profiles. The following table models the completion time at the receiver (yellow star) for each of the four cases in each profile. We use $\alpha$ to denote the latency ($\frac{1}{2}$ RTT), $\beta$ to denote the inverse bandwidth (time per Byte), and s the message size. The send is posted at $t_s$ and the receive at $t_r$; in the expected case, $t_s\geq t_r-\alpha$ and in the unexpected case, $t_s<t_r-\alpha$. The table shows the latencies for the worst-case $t_s$ and $t_r$ as discussed above. We assume that only headers can be buffered at the receiver.

\definecolor{lightgray}{rgb}{0.95,0.95,0.95}
\vspace{0.5em}
\noindent
\begin{tabular}{>{\columncolor{lightgray}}l|c|c|c}
	\rowcolor{lightgray} & Rendezvous (HPC)\footnotemark[1] & Deferrable Send (AI Full)\footnotemark[2] & Receiver Initiated (AI Min)\footnotemark[3] \\
	\hline\hline
	\cellcolor{lightgray} Expected & $t_s+\alpha+\beta s$ & $t_s+\alpha+\beta s$ & $t_s+3\alpha+\beta s$ \\
	\hline
	\cellcolor{lightgray} Unexpected & $t_r+\alpha+\beta s$ & $t_r+\alpha+\beta s$ & $t_r+2\alpha+\beta s$ \\
\end{tabular}
\vspace{0.5em}

\footnotetext[1]{If the send window is constant and exact, $\alpha$ of the second (read) step is overlapped with the incoming eager data.}
\footnotetext[2]{This mode can react to changes in the send window during the sending process optimizing the unexpected case.}
\footnotetext[3]{Assuming $t_s=t_r-\alpha$.}

Note that Rendezvous and Deferrable Sends are seemingly identical in runtime – this only applies to the case where the send window is constant and exact. Receiver-initiated transmission is expected to perform best for very large messages due to the additional RTT ($2\alpha$) penalty, which can be avoided if receives can be posted at $t_r<t_s-\alpha$. 

\subsubsection{RMA Read/Write}

RMA read and write are supported for workloads such as storage or one-sided distributed algorithms and synchronization (e.g., in graph problems~\cite{pushpull}). UE’s write is straightforward in that the full address, including Job ID, PIDonFEP, resource index, a target memory key, destination address, and offset is encoded into \emph{each} packet such that packets can be written out of order into the destination buffer. As a side note, this also applies to send/receive messaging.

UE’s read implementation aims to minimize the state at the target FEP for a typical read and to provide some fairness among readers. Thus, it utilizes \emph{single-packet reads} where the initiator issues a series of $\leq$ MTU-sized read requests that are satisfied at the target one by one (potentially out of order as well). If the target does not need completions, then only the initiator keeps a per-message state in this protocol. The initiator can rate-limit the requests if necessary, and many reads from many initiators can be interleaved at the target. This creates a complex relationship with the window-based flow control at the outgoing congestion-control context at the target that needs to be managed by each implementation. 

UE is designed to use two traffic classes (TCs) in the fabric.  The congestion management subsystem (CMS) leverages one TC for bulk data and a second TC for control traffic (e.g., ACKs).  This partitioning expedites the delivery of congestion information and reduces the loss of ACKs that carry that information. Although UE was primarily designed for best effort (lossy) networks, it also supports lossless networks, such as Slingshot~\cite{slingshot}. UE defines a separate mapping onto two traffic classes for lossless networks to prevent protocol-dependent (or message-dependent) deadlock.  Some of the first MPP designs~\cite{t3d,CrayT3D_ArchOverview} avoided a message-dependent deadlock using two message classes (or virtual channel classes). For message passing systems, it was formalized by Song et al.~\cite{cite-message-deadlock} based on the same principles as early deadlock-avoiding routing algorithms~\cite{cite-dally}. 

In a lossless network, a deadlock can arise when there is a resource dependency between a request and a response, such as between a packet and an ACK or a read and a read response. A deadlock arises when buffering is exhausted, and a target can no longer accept requests because it can no longer generate responses. Lossy networks resolve the deadlock by simply dropping packets if forward progress stops. Purely send- or write-based protocols can solve the deadlock utilizing cumulative acknowledgments, because it allows writes to continue being absorbed indefinitely without generating an acknowledgment. Protocols that incorporate reads and use lossless networks come in two forms: those that utilize two TCs (one for the read request and the other for the read response), and those that pre-reserve buffering for read requests at the target (like RoCE and InfiniBand). The requirement for pre-reservation of resources has long led to implementations of RDMA Read with poor performance, and those pre-reserved resources were typically associated with a connection construct (e.g., a queue pair) that is not present in UE; thus, we chose to utilize a second TC.

\subsection{Transport Packet Delivery Subsystem (PDS)}

UE introduces an innovative packet delivery subsystem that uses ephemeral Packet Delivery Contexts (PDCs) to manage the reliable transmission of packets from source to destination. PDCs can be established without incurring additional latency at first packet arrival. UE also prohibits fragmentation in the network and sends all but the last packet of each message with a full MTU payload. This significantly simplifies the packet tracking logic. 

UET’s PDS layer defines three packet types: \textbf{request packets}\footnote{Not to be confused with SES semantic requests and responses} send data (from initiator to target for write and send, or from target to initiator for read), \textbf{acknowledgment packets} carry information about previous request packets, and \textbf{control packets} are used for transport-specific controls (e.g., check the status of a packet at the receiver or the state of a path).

\subsubsection{Packet Transport Modes}

UE is designed to support a wide variety of use cases, including backward compatible workloads requiring strict ordering, completely out-of-order use cases, or anything in between. It provides four different packet ordering and reliability transport modes: \textbf{Reliable Unordered Delivery (RUD), Reliable Ordered Delivery (ROD), Unreliable Unordered Delivery (UUD), and Reliable Unordered Delivery for Idempotent Operations (RUDI)}. It does not support Unreliable Ordered Delivery because no use case could be identified at the time of specification. 

\textbf{Reliable Unordered Delivery (RUD)} is the default bulk transport mode. It should be used for any large-message transmission and for all transmissions if no wildcard matching is required (as in the AI profiles). A possible implementation that guarantees in-order message matching for RUD for AI Full could assign a message id (mid) to each posted send and receive (starting from zero). Then, the mid could be part of the matching (together with a communicator tag) and thus enforce that the first send to a specific target matches the first receive at that target, irrespective of the packet ordering on the wire. Each posted send or receive would increase that message id locally. RUD is considered the most efficient reliable transport mode UE offers because it enables packet spraying (see Section~\ref{sec:cms}). 

\textbf{Reliable Ordered Delivery (ROD)} can be used if strict in-order guarantees, such as those in RoCE or InfiniBand, are necessary at the packet level. This is, for example, needed for wildcard matching with in-order guarantees as required by MPI in the HPC profile. Here, a simple protocol could send all the eager portions of rendezvous messages over a ROD channel and issue the read portion over a RUD channel. Advanced message matching strategies may be able to relax this further~\cite{exampi-matching}. ROD may also be needed to communicate with minimally resourced endpoints such as in-network compute switches or simple storage appliances because it only requires implementing the simple go-back-N scheme, similar to RoCE. Yet, since ROD uses only a single network path per flowlet~\cite{cite-flowlet, flowcut}, it is considered the much less efficient reliable protocol in UE. 

\textbf{Unreliable Unordered Delivery (UUD)} can be used if unreliable datagram communication is desired. This can be useful in enabling software-based protocol implementations or system management tasks. 

\textbf{Reliable Unordered Delivery for Idempotent Operations (RUDI)} is intended to support idempotent operations that can be applied several times at the destination. For example, the same read or write can be applied as many times as necessary and does not change the outcome if the application does not access or change the value in between. This mode allows implementors to omit filtering duplicate packets/messages (retransmissions) at the target to improve efficiency of the implementation. It is the most scalable option because it does not require receiver state. Its downsides are that it is tricky to use (users need to ensure consistency across synchronized epochs) and it does not apply to nonidempotent operations such as atomic addition. This protocol is intended for very special use cases and is only available in the HPC profile.

All unordered packet transport modes may lead to out-of-order data arrival in main memory, as the SES layer does not reorder packets.

\subsubsection{UE headers}


Each packet carries headers for the various sublayers, many of which are optional or configurable. This leads to a number of possible packet header configurations of varying size. As shown in Figure~\ref{fig:arch}, each packet carries a standard Ethernet header of 14B and a Frame Check Sequence (FCS) of 4B. UET runs over UDP/IP with the option to run directly over IP with a 4B entropy header replacing the 8B UDP. The PDS header is 12B (16B if using RCCC congestion control, cf.~Section~\ref{sec:cms}) for RUD and ROD, 8B for RUDI and 4B for UUD. An optional 4B trailing end-to-end CRC immediately before the FCS provides an integrity check over headers and data. The SES headers include a 44B header for standard operations, 32B for matching messages up to 8kiB, and a minimal 20B header for nonmatching messages. When security is required, the optional TSS security header is 12B (16B if using explicit source identifiers) before the PDS header plus a 16B ICV is added at the end of the packet, right before the FCS. The ICV is far stronger than the PDS CRC, which can be omitted if an ICV is used. An overview of UE's headers is also shown in Figure~\ref{fig:arch}.

UE offers several header options to provide opportunities for vendor-specific differentiation. For example, congestion control fields such as the number of received bytes at the target congestion control context in RCVD\_BYTES, the time it took the target to process a packet locally in SERVICE\_TIME, or a direct number of credits to adjust the send window size in the CREDIT field. UE also supports SACK offset and a 64-bit Bitmap as well as optional OOO\_COUNT to support out-of-order packet handling.  These fields enable better flow control and congestion management. Proprietary information may also be carried using dedicated encodings for vendor-specific extensions. UE dedicates a portion of the encoding space to vendor-specific extensions in PDS types, semantic opcodes, and acknowledgment state. 

\subsubsection{Dynamic PDC creation}

\begin{figure}[h!]
	\centering
    \includegraphics[width=\linewidth]{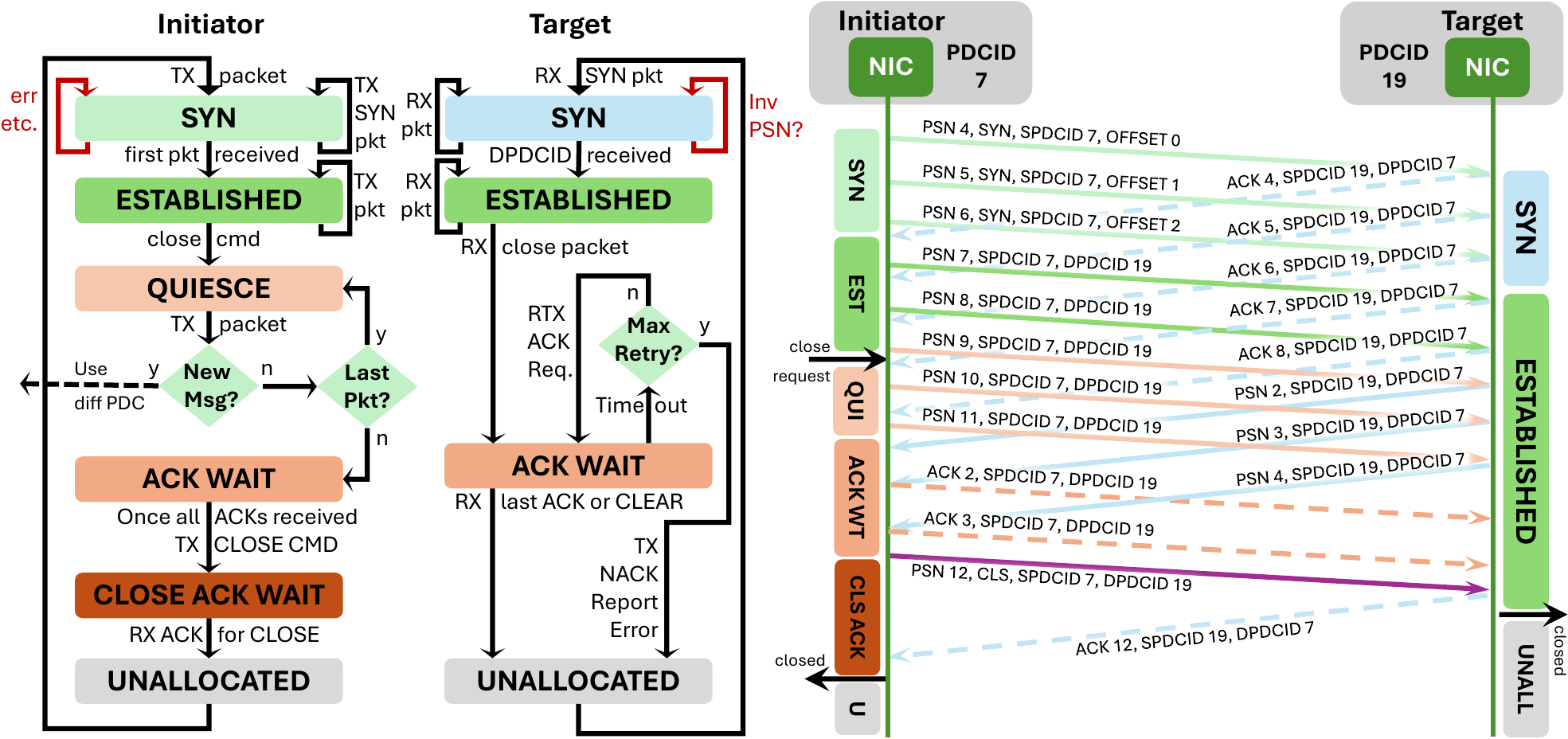}
	\caption{Ultra Ethernet Transport's Dynamic PDC Creation State Machine.}
	\label{fig:pdc}
\end{figure}

Packet delivery contexts can be created without incurring a latency penalty because all the states necessary to establish a connection are carried in the headers of the data packet. The protocol even supports connection establishment for out-of-order RUD channels. Figure~\ref{fig:pdc} illustrates a simplified PDC state machine for the initiator and target on the left side and a scenario where a PDC is established and closed on the right side. Packet Sequence Numbers (PSNs) are used to identify packets and PDC IDs are used to identify PDCs at either source or destination. The state machine excludes much of the error/packet drop handling during initialization (the red self-loops at the SYN state) and intermediate state transitions that may require retransmission. 

The scenario on the right illustrates three messages: a five-packet write (PSN 4-8) followed by a three-packet read (PSN 9-11 and PSN 2-4) during which a close request is received at the initiator. 

This scenario starts with no PDC and an SES request to write. The first packet transmission builds a PDC with ID 7 at the source and sends three packets starting with a random PSN 4 to the destination. When the first packet arrives at the destination, it creates a new PDC, in this case with ID 19. The newly established PDCID is sent back in each ACK packet. Once the source receives the first ACK packet with a valid target-assigned PDCID, it transitions to the established state and continues sending packets without SYN. Note that the source has been sending at full rate during PDC establishment! Once the first packet without SYN arrives at the destination, it transitions to the established state. With both sides established, the PDC will function with a synchronized PSN space as usual. 

The initiator always starts a PDC close when the PDC is idle. An initiator is not required to immediately close an idle PDC; the target can request that the initiator close a PDC using a control packet or flags in an ACK. Once the source begins closing a PDC, it enters the QUIESCE state, where it continues to send packets for messages that have started before the close, but refuses new messages. Once all messages have been drained, it enters the ACK WAIT state until it has received all outstanding replies. Once all replies have been received, it sends a final close command to the target, which closes the PDC.  Once the final ACK for the close command has been received at the source, the source PDC is freed.

\subsubsection{Fast Packet Loss Detection}

When a UE packet is dropped, it must be retransmitted from the source. Timeouts are used today as standard ways to retransmit dropped packets. However, a timeout is rarely a reliable detector of packet drops – a packet may time out at the source while it is still waiting inside a switch buffer, leading to duplicate transmission and wasted bandwidth. Guaranteeing that timed-out packets are not waiting in the network would require a very high time-out value (especially in deeply buffered datacenter switches and long paths). High timeout values will lead to very low bandwidth utilization of links with congestion drops. In general, choosing the right timeout to balance needless retransmissions and bandwidth utilization is incredibly difficult in practice. Thus, UE defines alternative, faster loss detection mechanisms.

We distinguish three kinds of packet drops (we introduce this here as “the three Cs in networking”): \textbf{Congestion drops}, where packets are dropped when switch buffers are full, \textbf{Corruption drops}, occur when packets with bit errors fail the checksum, and \textbf{Configuration drops}, when the network is configured to drop, e.g., with firewalls or time-to-live expiration. UE describes three optional loss detection mechanisms that can detect Congestion drops, one of them can also reliably detect Corruption drops.

\textbf{Packet trimming}~\cite{ndp,trimming} is the first and simplest scheme, but requires switch support. In this scheme, switches can be configured to trim the packet payload of data packets that would be dropped and forward the remaining headers, potentially on an elevated priority, to the destination. Upon receiving a trimmed packet, the destination knows that the payload was dropped and can quickly request a retransmission from the source. UE defines the detailed behavior of switch-based packet trimming. Trimming cannot detect Corruption drops. 

\textbf{Out-of-order count} can be used to estimate the number of packets lost by calculating the distance between the last received PSN and the earliest missing PSN. It can either be calculated at the destination and sent via the optional OOO\_COUNT ACK extension header field to the source or it can be estimated at the source. If this counter exceeds a certain threshold, then the system can assume that packets are lost. This scheme is more accurate than timeouts, but it may also send duplicate packets if sprayed packets have very different latencies across different paths. 

\textbf{EV-based} schemes can be used to precisely detect loss. The simplest scheme could keep an ordered list of (EV, PSN)-pairs that were sent at the source and match each incoming ACK against this list. Since UE ACKs are issued in the order of arrival and carry the same EV as the acknowledged packet, and an EV picks a unique path in the absence of hard failures, the source can simply look for the oldest entry in the send list with the same EV, and if the received PSN does not match, it can infer that all packets with the same EV and PSNs smaller than the received one were lost. This simple inefficient scheme can be optimized with the concept of k slots over which packets are sprayed round-robin. Now, the PSNs expected at each slot are i,i+k,i+2k,… and loss can be inferred if ACKs arrive for later than expected PSNs. Tail losses can be detected by sending probe control packets along specific paths (EVs). Such probe packets can also be used when the EV in a specific slot is to be changed for load balancing reasons.

\subsubsection{Packet Spraying and Reliability}

Packet-sprayed networks pose a unique challenge for reliable transmission and messaging semantics mechanisms because packets do not arrive in the order they were sent. The target maintains a bitmap for message completion and reliability management. UE defines a cumulative ACK (CACK) PSN such that an ACK packet can acknowledge multiple data packets. This mechanism is useful if ACK packets are lost. Furthermore, UE allows optional ACK coalescing to enable the receiver to not acknowledge every packet. Coalesced ACKs support systems for which it is too expensive to generate an ACK per packet. UE ACK packets also carry a 64-bit SACK bitmap, where a set bit indicates a packet arrived at the target that can be used to infer loss of some packets. 

Reliable, unordered protocols pose a specific implementation challenge: How out of order will the packets be? After all, the target side must track packets that arrive out of order as having been received, often using a structure such as a bitmap. Unfortunately, in the worst case, unordered networks may deliver the last packet first. Small messages on high-bandwidth networks with target round-trip times of $10 \mu s$ lead to a large bitmap to cover the bandwidth delay product (BDP) - perhaps larger than every implementation wants to support. Given that UE is connectionless such that communications with a peer can begin without a round-trip handshake, and given that bitmap resources might be dynamically allocated, UE included a concept known as the Maximum PSN Range.  The MP\_RANGE field of the wire protocol allows the target of a communication to dynamically adjust the allowed PSN range that is outstanding to it.  This explicitly limits the amount of packet tracking resources that must be used.  Like most things in UE, the expectation is that communications will start with an optimistic view: that the target can track whatever is sent. This optimizes performance for good implementations and well-tuned applications. Later, if the target is resource constrained, the protocol cleanly recovers and is able to prevent further resource overruns.

\subsection{Transport Congestion Management Subsystem (CMS)}\label{sec:cms}

UET’s Congestion Management is designed to only require minimal support from the switch infrastructure such that it can be deployed in legacy network settings. CMS entails both Congestion Control (CC) to limit the number of Bytes in the network and Load Balancing (LB) to select good paths. It only requires basic ECN marking on egress and ECMP for load balancing. UE’s CMS is designed and tested in simulation for best-effort networks. It can take full advantage of fast loss detection mechanisms, such as packet trimming. 

When used in a shared network with other traffic, UE assumes that it can be run in its own traffic class using the Class of Service (CoS) functionality of Ethernet. Support for advanced congestion signaling techniques and other in-network telemetry may be added in future versions. UE defines several control packet types to explore the state of paths that may be used to gather additional congestion information or detect loss, as we described before.

Fundamentally, UE’s congestion management is implemented in a Congestion Control Context (CCC) that serves one or more PDCs. Usually, a CCC coordinates all PDCs of the same traffic type between the same two FEPs. At the sender, a CCC maintains a congestion window that limits the number of unacknowledged bytes in the network. It can be seen as the number of Bytes (credits) that a sender can send into the network before it needs to wait for them to be acknowledged. 

UE offers two complementary congestion control algorithms: Network Signal-based Congestion Control (NSCC) and Receiver Credit-based Congestion Control (RCCC). NSCC runs a control loop at the source to adjust window sizes and is available on each UE NIC. RCCC defines a receiver-driven credit-assignment algorithm that is optional to implement. Either of the two can be disabled at runtime, and thus a system that implements RCCC can run all possible setups, NSCC alone, RCCC alone, or both in tandem.

\subsubsection{Network Signal-based Congestion Control (NSCC)}

NSCC is inspired by previous algorithms and an early version has been documented and analyzed by UE’s CMS team~\cite{smartt} and later, a subgroup published a related analysis~\cite{strack}. It combines two fundamental signals: round-trip time (RTT) and ECN marking from the request to the response packet. In the following, we assume per-packet ACKs where both can be measured precisely. Both Cumulative ACK (CACK) and Selective ACK (SACK), which may acknowledge multiple packets, are supported but may result in more complex cases. 

ECN marking~\cite{rfc-3168} can be seen as a statistical single-bit signal. Switches are configured to mark packets probabilistically such that if enough packets are observed and the state of the switch buffers along the path does not change, the source will get an accurate picture of the path. Yet, if multiple switches along the path would set the ECN Congestion Experienced (ECN-CE) flag (short “ECN bit”), it is still only set once. Unfortunately, this method may need many packets to converge, and the state of the buffers usually changes constantly. UE requires ECN to be marked at egress (when packets are leaving the switch), which is different from RFC 3168 but common practice in switches today. With egress marking, the signal generation skips the queue and thus propagates quickly back to the sender. Several existing CC schemes rely solely on ECN~\cite{dcqcn}.

RTT on the other hand can be seen as a multi-bit signal because it can record the exact time from request to response, depending on the resolution of the local timers. However, RTT significantly delays the congestion signal because it cannot skip any congested switch queues. If switch buffers are large and well filled and a path has many hops, the time before any RTT information is available for CC at the sender can be substantial. Even in a steady state, the RTT information may quickly become outdated as the state of the network changes rapidly. UE RTT measurement should exclude the service time at the receiver. Several CC schemes are based solely on RTT~\cite{timely}.

NSCC uses both signals, the quick ECN 1-bit signal and the lagging RTT multi-bit signal, and defines the following four cases when an ACK arrives:
\begin{enumerate}
\item	ECN mark + low RTT indicates that congestion is building up at least at one switch buffer
\item	ECN mark + high RTT indicates that the network is congested, may be overloaded
\item	No ECN mark + low RTT indicates that the network is not congested, may be underloaded 
\item	No ECN mark + high RTT indicates that the network congestion is going down
\end{enumerate}

To classify low and high for RTT, NSCC maintains a value representing its best guess of the expected unloaded RTT at any moment. NSCC does not react in case (1) but adjusts the window size in all other cases. In case (2), it decreases the window size aggressively for each incoming packet. In case (3), it quickly increases the congestion window with a guess based on the measured and expected RTT. In case (4), NSCC gently increases the congestion window.

In addition to those four cases, NSCC uses the Quick Adapt (QA) algorithm to estimate bottlenecks in the network (e.g., incast at the receiver or congestion at a switch in the network). It utilizes the packet loss signal (e.g., fast loss detection such as trimming) to quickly adjust the rate to the proportion of successfully delivered packets. Detailed explanations, measurements, and intuition behind the four cases above and the QA algorithm can be found in the SMaRTT paper~\cite{smartt}.

\subsubsection{Receiver Credit-based Congestion Control (RCCC)}

UET’s RCCC is inspired by earlier works using such schemes~\cite{ndp}. Unlike in NSCC, the sender does not adjust the window size based on network signals. Instead, it reduces the window size when sending and increases it when receiving credits from the receiver. All credit management is handled by the receiver, who knows the exact number of incoming flows and supplies credits to the senders at a certain rate. This can be highly beneficial for several traffic patterns. RCCC can also consider the sources’ demands and schedule per-source rates for workloads with irregular demand effectively. 

One main strength of this approach lies in its simplicity and efficacy for many relevant workloads. RCCC can control traffic well in the event of receiver congestion (incast), which NSCC can only guess with QA. However, it cannot react to congestion in the network or outcast at the sources. Thus, UE recommends enabling NSCC in oversubscribed networks as well. RCCC can also use proprietary implementations that use other signals at the destination to adjust the rate. Both algorithms can be run jointly to achieve the highest throughput. In the following, we will give an overview of three congestion patterns and how each algorithm may perform.

\subsubsection{Incast, Outcast, and Oversubscription}

Both RCCC and NSCC are optimistic and start at full rate (line rate) with a window that is at or slightly larger than the BDP. They will continue to run at that rate if there is no network congestion. The main purpose of both schemes is to control the network load in congestion situations. 

\begin{figure}[h!]
	\centering
    \includegraphics[width=\linewidth]{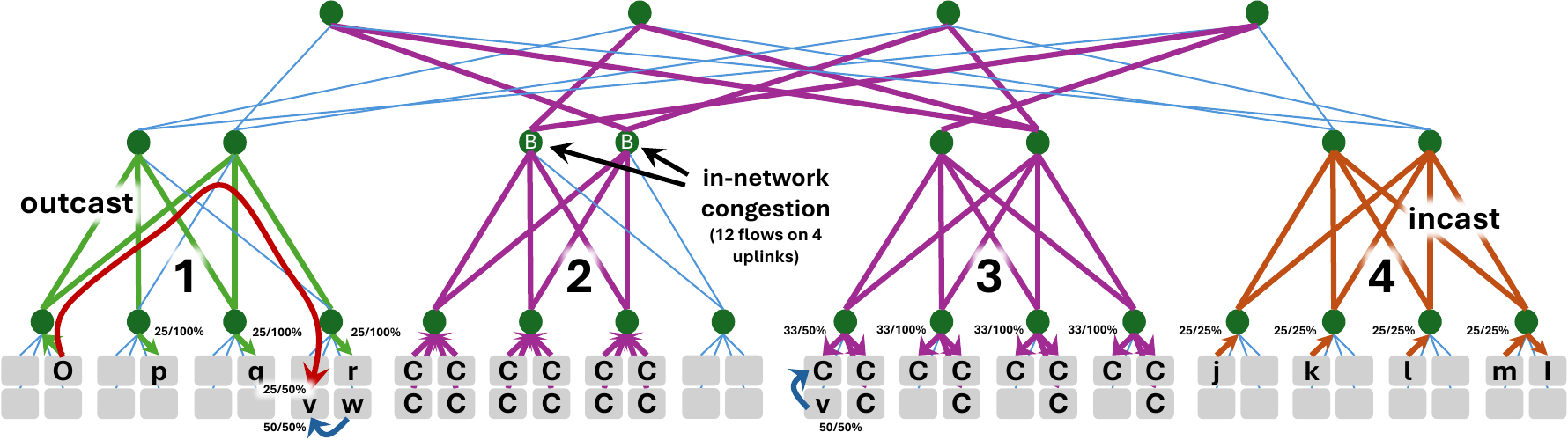}
	\caption{Ultra Ethernet Transport Congestion Examples.}
	\label{fig:cong}
\end{figure}

Figure~\ref{fig:cong} illustrates three common congestion scenarios: Flows from source to destination are shown as colored edges in the 2:1 oversubscribed fat tree. The right group 4 shows a typical incast scenario in which nodes j, k, l, and m send to node I. RCCC detects all incoming flows and assigns 25\% of the bandwidth to each of them. The notation “x/y\%” at each flow means that RCCC assigned y\% of the flow and the flow delivers x\%. For incast, the bandwidth utilization is optimal. NSCC would converge to the same bandwidth utilization, but would require more time (either through drop detection with QA or the RTT/ECN-based mechanisms). 

The opposite situation, called outcast, that is trivially handled by NSCC but may be problematic for RCCC is shown on the left side in group 1. Here, node O sends to nodes p, q, r, and v. Furthermore, node w also sends to node v. RCCC assigns 100\% to incoming flows at p, q, and r and 50\% to incoming flows at v, given the local incast. Now, node O can only send at 100\% and thus will send at 25\% to each of these nodes, which is not problematic per se. Yet, RCCC running on node v only assigns 50\% to the flow from node w but could have assigned 75\%, thus losing 25\% of the bandwidth. NSCC would again converge slower, but eventually to the optimal configuration.

A more complex in-network congestion scenario is shown in the middle groups 2 and 3. Here, 12 nodes marked with C in group 2 send each a line rate flow 1:1 to the corresponding nodes in group 3, also marked as C. In a nonblocking network, this would be possible, yet, in our oversubscribed network, the bottleneck switches (marked B) receive 12 flows but only have 4 uplinks; thus, they can only satisfy 33\% of the bandwidth. Now consider that node v in group 3 sends to node C as shown. RCCC on node C would only assign 50\% of the bandwidth to this flow even though it could assign 67\%, again leading to inefficient bandwidth utilization.

Outcast, incast, and in-network congestion are typical scenarios that motivate the use of NSCC even though it may converge very slowly due to the signal propagation delay. RCCC, while having a perfect view for incast, may benefit from NSCC in two scenarios, and thus UE recommends enabling NSCC if RCCC is enabled. There may still be cases, e.g., full-bandwidth networks with workloads that do not exhibit outcast patterns where the simple RCCC algorithm is expected to perform well in isolation. 

\subsubsection{Destination Flow Control}

Regulating network rates through congestion control is only part of the puzzle for successful data transmission between endpoints. Slow or overloaded receivers, destination memories, or local buses may not always be able to service the full rate of incoming packets. This would lead to drops between the destination FEP and destination memory and require retransmissions. UE’s Destination Flow Control (DFC) is a simple protocol to throttle the sender even beyond what the network can deliver. 

In RCCC, DFC works at the receiver by simply reducing the credit rate to the sender and thus adjusting the sending rate to destination congestion. For NSCC, the receiver sends an 8B congestion window penalty value that the sender uses to scale to change to the congestion window. DFC enables both algorithms to react quickly to situations of transient or steady destination memory congestion.

\subsubsection{Load Balancing}

UE defines various mechanisms for load balancing for RUD and RUDI traffic. All ROD traffic for a given PDC must be routed along the same path in UE 1.0. UE does not allow an endpoint to select specific paths, but it can use EVs to control when to change a path. The actual path is determined by the hash function applied to the header, including the EV. It is possible to build a system that only uses the EV as input to special hash functions that allow endpoints to select a specific path, but it may be complex to deal with failures in such a setup. UE only assumes minimal guarantees in the absence of failures: (1) the same EV will select the same path, and (2) it is likely that a different EV will choose a different path. This is usually achieved by well-mixing hash-functions in use today. 

UE’s transport is designed to function well even if each packet chooses a different EV/path. Thus, load balancing can be performed independently of other functions, such as reliability or congestion control. Note that UE’s congestion control operates on a CCC-to-CCC basis and not on a per-path basis. Thus, interactions between LB and CC must be considered in an overall system architecture. The load balancing strategies are implemented at the CCC level and apply to all associated PDCs.

The simplest LB scheme is oblivious spraying, where the source CCC changes the EV for each injected packet. The number of entropies may be limited, or they may be chosen randomly, yet oblivious spraying does not take path feedback into account. UE recommends using oblivious spraying together with a fast loss detection mechanism such as trimming. In this way, lost packets are quickly recovered and the network can be statistically well balanced. Oblivious spraying works best if all endpoints in the same traffic class are configured for it and no other traffic interferes.

If other traffic, for example, ROD traffic, shares a network or traffic class, oblivious spraying may not work well. In these scenarios, UE recommends path-aware spraying that tracks the throughput for different EVs and balances the load accordingly. LB schemes detect if an EV is congested if either trimmed packet NACKs or ECN-CE-marked ACKs arrive. UE also offers a mechanism to distinguish trims on the last hop from trims in the network because changing the EV for last-hop trims would not be effective. UE generally leaves it up to the implementation to select EVs and how to track the quality of paths. It provides two example schemes for such selection mechanisms. 

(1) Recycled Entropies Spraying (REPS)~\cite{reps}, does not keep any endpoint state in its minimal form. REPS relies on a simple control loop: When sending, it uses EVs that were returned via ACKs. If there are no “recycled” EVs available (e.g., at startup), it will pick new random EVs. Those EVs will then be used for sending new packets, and then they are returned through ACKs. This mechanism is self-clocking, which means that if there is one slow and one fast path, for example, eventually both paths will be used at their optimal capacity due to the rate of returning ACKs. Thus, in steady state and for unchanged conditions, REPS will converge to the optimum bandwidth along each path. 

(2) Another scheme in the UE specification proposes to keep a set of random EVs and an associated bitmap at the source. The source sets a bit if the corresponding EV detected congestion. The source would then rotate through EVs when sending packets and skip the EVs in the rotation for which the bit is set. A skipped bit is unset in the bitmap so that it will be selected during the next round. The number of EVs can be dynamically adapted to adjust the rate at which congested EVs are retried~\cite{strack}. 

All LB schemes are optional in UE and vendors can invent their own scheme in addition. Different LB schemes may interfere in non-trivial ways in the network, and care should be taken when designing systems.

\subsection{Transport Security Subsystem (TSS)}\label{sec:tss}

UE is designed from the ground up with security in mind. Its design was inspired by sRDMA~\cite{srdma} and ReDMArk~\cite{redmark} as well as by the PSP~\cite{google2022psp}, IPSec~\cite{Atkinson1995IPsecArch}, and MACSec~\cite{IEEE8021AE-2018} specifications. TSS provides end-to-end confidential and authentication services between a group of FEPs using a 'zero trust' security model.  As a transport security service, it does not define interaction with the host memory. However, it is designed to work in conjunction with other mechanisms such as the PCI Express Trusted Execution Environment Device Interface Security Protocol (TDISP) to provide a complete security solution. The SES, PDS, and TSS layers support several counters that record errors that may indicate attacks.

\subsubsection{TSS Encryption}

TSS is an optional service that authenticates the PDS headers, SES headers, and payload along with the IP address of the packet. Flexibility is provided to configure which parts of the packet are encrypted as a trade-off between information leakage and network visibility.  In a zero-trust model, the switches are untrusted but need to provide congestion and quality of service information. Specifically, ECN congestion marking and packet trimming are not authenticated and are carefully handled to minimize the attack surface. For example, trimmed packets must not trigger PDC creation. 

TSS defines the concept of a secure domain (SD), which provides confidentiality and authentication services between a set of FEPs, achieving full data isolation to other SDs and unencrypted traffic.  An SD can be used by multiple jobs or isolated to a single job to provide a scalable architecture. All members of an SD use the same symmetric cryptographic secure domain key (SDK) as the basis for authentication and confidential services. This key is distributed by a management entity (SDME) to all members of the SD. The SDME is also responsible for updating the key and domain membership.

Depending on the use case, the SD provides several mechanisms to obtain the symmetric keys used for each packet. The simplest approach uses the SDK directly. A second mode, optimized for distributed communication, uses a Key Derivation Function (KDF), a deterministic and non-invertible function, which derives a new key from the domain key and some arguments (e.g., source or destination address) to obtain a source key. Finally, a mode for client-server communication uses KDF and the destination addressing to derive a per source key which improves server scaling.  

UE uses Authenticated Encryption with Associated Data (AEAD). The default algorithm is AES-GCM~\cite{aes-gcm} with a 256b key and a 16B authentication tag or ICV (integrity check value).  A critical security parameter is the initialization vector (IV) of the cipher, which is treated as nonce (number used once) and must not be reused across all members of the group. One of the key properties of an SD is that all members of the domain have the same key, and ensuring that the IV is not reused is a key part of the protocol design. A Time Stamp Counter (TSC) is carried in each packet and consists of a 16 bit epoch identifier and a 48 bit packet counter. The epoch is managed by the SDME to facilitate the addition and removal of members, and the packet counter is incremented for each packet sent by a FEP. 

To address attacks on AES-GCM using a fixed nonce, authenticated data, and encrypted data as outlined in IETF RFC 9001 Appendix B, TSS defines an IV mask which is XORed to the packet IV before it is given to the cipher. This ensures that the IV is randomized and not directly accessible to an attacker. The IV mask is stored as SD or derived using a KDF function.

Several mechanisms are provided to manage key lifetime, which is this context is the number of packets encrypted with the same key. Depending on the assumptions of the security model (single vs. multi-user), the limit is between $2^{27}$ and $2^{34.5}$ packets.

The key rotation changes the SDK and the Association Number (AN). Each receiver maintains multiple AN keys per SD to support seamless key rotation. Optionally, portions of the TSC can be used in a KDF to periodically change the source key 'automatically' as the packet count increases. 

\subsubsection{Preventing Replay Attacks}

To prevent replay attacks when PSNs wrap, each PDC using encryption must be closed and re-opened after sending two billion packets. Furthermore, the PDS is fundamentally stateless, thus, it may be susceptible to replay attacks when establishing a new PDC if it would accept any initial packet sequence number (PSN). The transactions carried in the initial packets could modify the target’s state in unwanted ways. Thus, when using encryption, the PDS establishes initial PSNs securely using one of two ways.

The first way uses an additional round trip time to establish a secure PSN. The first packet sent from the initiator to the target is a NOP to query the starting PSN. The target then replies with a locally known random starting PSN and creates a PDC in pending state. Although this scheme to agree on a random starting PSN prevents replay attacks, the additional RTT may not be desirable.

The second scheme supports zero-overhead startups but requires some coordination within the SD. Here, two values are stored associated with each SD: start\_psn and expected\_psn, both initially zero. The start\_psn is used for new requested PDCs and the expected\_psn is the minimum accepted PSN for new incoming PDCs. If a target receives a connection request with a PSN smaller than expected\_psn, then a NACK is generated indication a starting PSN to be used, otherwise, it accepts the request. When a PDC is closed, expected\_psn at the target is increased to the current PSN+1 of the closing PDC. This ensures that no replay attack can be performed. The new expected\_psn is sent back to the source as part of the close PDC ACK. The source then updates start\_psn to this value to ensure that future connections are zero-RTT. This scheme is not guaranteed to always lead to zero-RTT PDC establishment and falls back to one-RTT establishment if it fails. It allows for additional heuristics to set start\_psn or expected\_psn to improve the probability. Both schemes interoperate. 

\subsection{Link Layer Features}
\label{sec:ll}

The Ethernet-compatible Link Layer introduces two independent optional features, Link Level Retry (LLR) and Credit-Based Flow Control (CBFC). To retain Ethernet compatibility, the Link Layer Discovery Protocol (LLDP) is used to negotiate with peers to enable LLR and CBFC.

\subsubsection{Link Level Retry (LLR)}

With increasing link bandwidths and potentially growing error rates, latency-sensitive workloads can be negatively impacted by the traditional end-to-end retransmission. LLR handles errors locally at the link level when transmitting Ethernet frames. The transmitter stores all LLR-eligible frames in a replay buffer and assigns a sequence number encoded in the preamble. The receiver link sends acknowledgments back to the source once frames have been received to free the replay buffer. Recovery of last packets uses go-back-N retransmission, where missing sequence numbers are detected based on the arrival order and send negative acknowledgments (NACKs). Furthermore, retransmission timeouts are used to prevent losing the last packet (tail loss) or NACKs. Control packets such as (N)ACKs are sent via the Physical Layer’s PCS subsystem encoded as 8B Control Ordered Sets. 

In the context of UET's new protocol design that was partially inspired by the need to eliminate retransmit timeouts and go-back-N retransmission, these choices may seem odd on the surface; however, the link level environment is substantially different than a transport that traverses a system at scale.  In the link-level environment, the retransmission timeout can be tightly bounded by the round-trip time of the link ($\approx$1 microsecond), and packets are only discarded due to physical layer errors (e.g., uncorrectable FEC symbols or FCS errors).  Congestion plays no role in packet discards or in round-trip time; thus, a go-back-N protocol augmented with a short retransmit timeout is sufficient and prevents the link level from needing to understand any ordering requirements that might exist for the traffic. 

\subsubsection{Credit Based Flow Control (CBFC)}

UE’s CBFC provides link-level flow control and can be used in conjunction with or in place of Converged Ethernet’s Priority-Based Flow Control (PFC). Both schemes are intended to provide a nearly lossless packet service, eliminating congestion packet drops which are caused by a lack of buffering for the arriving packet. Such congestion drops can be the main cause of network performance degradation in end-to-end reliable systems that use a go-back-N retransmission strategy, such as UE’s Transport ROD service. Even with optimal congestion management, packet drops can occur due to switch buffer overflow, extended network paths, and highly variable traffic patterns causing at least one additional RTT latency to resolve using an E2E retransmission protocol. CBFC or PFC can optimize these scenarios by creating a lossless Link Layer (Layer 2) fabric.

CBFC offers several advantages over PFC:
\begin{enumerate}
\item	It requires less buffer space for guaranteed delivery.
\item	Senders can leverage credit availability information for each virtual channel for local scheduling.
\item	It is simpler to configure for guaranteed delivery.
\item	It can use less buffer space for virtual channels that require limited throughput. 
\end{enumerate}

PFC requires an RTT+MTU headroom buffer for each port for lossless operation. There are several ways to optimize the headroom buffer from a switch perspective. However, CBFC allows for more efficient allocation and management of this buffer; see Hoefler et al.~\cite{roce-issues} for more details.  

CBFC uses two 20-bit cyclic counters at both the sender and the receiver to track credits consumed and credits freed at the receiver based on buffer occupancy. These counters are maintained per virtual channel and are periodically synchronized to avoid credit loss. UE CBFC defines two message types, a high-frequency update message from receiver to sender encoded as 8B Control Ordered Set and a low-frequency update message from sender to receiver encoded as 64B packet.

Before sending a packet, the sender checks credit availability locally, and once a packet is sent, deducts the packet size from the credit. The receiver gives credit back to the sender using update messages once a packet leaves its buffer. The sender periodically updates the receiver with update messages to prevent credit loss. In general, the protocol will guarantee that packets are only sent if there is enough buffer at the receiver.

\section{Summary and Conclusions}

UE has been designed in a 30-month sprint by many people. Its first version innovates on many fronts and enables the Ethernet ecosystem to deliver efficient, cost-effective, and secure networking for AI and HPC workloads. It would be very surprising if we would not require several rounds of errata clarifications and fixes to the first specification. The first products have already been announced and should be available within months.

\subsection*{Acknowledgments}

The authors thank the whole Ultra Ethernet Consortium and many individuals for their contributions. TH especially thanks his peers in the circle of editors who collaboratively produced most of the Ultra Ethernet Transport text, Karen, Eric, and Keith for contributing to such a constructive and productive environment amid all the challenges of defining a specification with 100+ member companies and 1000+ members subscribed to the transport working group mailing lists. The authors thank Mikhail Khalilov, Marcin Copik, Marcin Chrapek, Daniele De Sensi, and Tommaso Bonato for comments that helped improve the manuscript.

\bibliographystyle{ACM-Reference-Format}
\bibliography{uec-authors-cut}

\end{document}